# What do university rankings by fields rank? Exploring discrepancies between the organizational structure of universities and bibliometric classifications

Nicolás Robinson-García[1], and Clara Calero-Medina[2]

[1] EC3: Evaluación de la Ciencia y de la Comunicación Científica, Departamento de Información y Comunicación, Universidad de Granada, 18071 Granada, Spain. E-mail: elrobin@ugr.es

[2] Centre for Science and Technology Studies, Leiden University, The Netherlands. E-mail: clara@cwts.leidenuniv.nl

**Abstract**

University rankings by fields are usually based on the research output of universities. However, research managers and rankings consumers expect to see in such fields a reflection of the structure of their own organizational institution. In this study we address such misinterpretation by developing the research profile of the organizational units of two Spanish universities: University of Granada and Pompeu Fabra University. We use two classification systems, the subject categories offered by Thomson Scientific which are commonly used on bibliometric studies, and the 37 disciplines displayed by the Spanish I-UGR Rankings which are constructed from an aggregation of the former. We also describe in detail problems encountered when working with address data from a top down approach and we show differences between universities structures derived from the interdisciplinary organizational forms of new managerialism at universities. We conclude by highlighting that rankings by fields should clearly state the methodology for the construction of such fields. We indicate that the construction of research profiles may be a good solution for universities for finding out levels of discrepancy between organizational units and subject fields.

-------------------------------------------------------

# Introduction

One of the most common approaches in bibliometrics for benchmarking multidisciplinary entities such as universities, research teams or institutes, is the use of classification-based tools and indicators (i.e., Moed, 2010; Leydesdorff & Ophtof, 2010; Waltman & van Eck, 2012b). This is the case in university rankings, which are now incorporating league tables by fields as a response to criticisms due to an over-simplistic perspective as these rankings tend to reduce the complex framework of universities' activity to a single dimension (see e.g., van Raan, 2005). The first one to include disciplinary-oriented league tables was the Shanghai Ranking, launching in 2007 rankings by five broad fields and in 2009 five more rankings in specific disciplines. Since then, many other international rankings have followed such perspective, such as the Times Higher Education World University Rankings, the QS Rankings or the National Taiwan University Rankings, for instance. The Leiden Ranking has been the last one to follow this trend, including in its 2013 edition rankings by five broad areas. Others, such as the Scimago Institutions Rankings do not show league tables by fields but include a specialization index.

This is partly because of the influence disciplinary specialization may have on research evaluation (López-Illescas, Moya-Anegón, Moed, 2011) which means that one must identify universities with similar disciplinary focuses (García, Rodriguez-Sanchez, Fdez-Valdivia et al., 2012) as an aid to interpret such comparisons. Also, global comparisons may be 'unfair' to certain types of universities as their subject profile may influence their positioning (Bornmann, Moya-Anegón & Mutz, 2013). Cheng & Liu (2006) already attempted at identifying disciplinary-oriented institutions by using clustering methods and, in a more recent study, Bornmann et al. (2013) developed a web application which maps centers of excellence according to different fields. All these evidences show the need to bypass the use of global rankings and focus on developing field-based tools.

The most commonly used classification system in bibliometrics is the one designed by Thomson Scientific (TS), which groups scientific journals following heuristic criteria based on citation data (Pudovkin & Garfield, 2002). Although it shows some limitations when used for bibliometric purposes (Glänzel & Schubert, 2003; Waltman & van Eck, 2012a), it seems to be a practical and plausible way to aggregate categories into areas when developing rankings by fields (see e.g., the 2013 edition of the Leiden Ranking which now includes rankings by five broad fields or the Spanish fields-based I-UGR Rankings described in Robinson-García et al., 2013a based on the indicator developed by Torres-Salinas et al., 2011). But this approach based on universities' output seems counter-intuitive when being read by ranking consumers, as they expect to see a *bottom-up* methodology which would determine the institutional structure of universities and hence, develop league tables according to their units (faculties, departments, etc.). Such granularity in the information provided by rankings has already been suggested elsewhere (Bornmann, Mutz & Daniel, 2013) as there are significant differences in terms of research performance between research units of the same institution. This would allow an attribution of the performance of a given university in a particular field to researchers assigned to the units related to such field. However, this is not always possible as there is no unified database containing such information, meaning that universities should have to be involved in the data collection process (van Leeuwen, 2007). Because of this, university rankings usually adopt a *top-down* approach, a reasonable solution but one which usually leads to misinterpretations by media and research policy makers.

Even so, one could suggest the use of the address data included in publications when constructing rankings by fields. De Bruin & Moed (1993) already suggested working with address data in order to develop a subject classification scheme based on an institutional structure. They departed from three basic assumptions: 1) scientific activity can be analyzed in terms of collaboration between research groups, 2) organizational units reflect to some extent the scientific scope of their members and 3) researchers indicate in their publications the organizational units in which they work. For this, they created a genealogical structure of the address data in order to identify cognitive terms from those which weren't and then applied a clustering method to isolate each sub-field. However, address data presents many problems as the authors acknowledged, seeming unfeasible to do this at a large scale due to the heterogeneity of universities' structure, the possible changes over time as a result of organizational re-shuffling, and the growing complexity of the problem.

Finally there is another aspect that should be mentioned. Although desirable, any attempt to develop more accurate and precise bibliometric rankings can bring other unexpected issues. The bibliometric field has an applied nature and empirical roots when compared with more basic sciences. This is why it does not only require detailed and full characterization of the analyzed entities, but also a certain level of assertiveness and security over its applicability in real situations; leaving a threshold of uncertainty. Hence, although a solution to a given problem may be theoretically correct, it may be wrongly interpreted or adjusted in a certain context. Such tension between the accuracy and precision demanded by any scientific tool, and the security on any statement needed when facing possible research policy applications is explicitly defined in Duhem's Law of Cognitive Complementarity. In it, Duhem highlights the inverse relation between detail and security, stating that in order to access to truth one needs certain levels of vagueness which will secure its reliability in any given situation (Rescher, 2006). Any attempt to create rankings by fields exposes itself to such dilemma, as the applicability of such fields on the organizational structure of universities may differ from one to another.

## Objectives of the study

This paper highlights the difficulties university research managers and other ranking consumers may have when understanding university rankings by fields as they misinterpret them by expecting to see in those fields a reflection of the structure of their university. As university rankings producers, bibliometricians must not only be transparent on the methodology and data employed, but also ensure a reasonable interpretation of the results they offer. We examine the relation between the institutional organization of science as reflected by authors' affiliation data and their research output. Specifically, our aims can be resumed in the following research questions (RQ):

RQ1. Do rankings by fields represent the structure of universities? De Bruin & Moed (1993), suggest that address data may be useful for identifying scientific fields and domains. Is there some kind of correspondence between the fields constructed by rankings and smaller organizational units such as departments, research group, faculties, etc.?

RQ2. Can we provide research policy managers and ranking consumers with indicators that they can use in order to understand to which degree each field (based on the Web of Science subject categories) corresponds with the output of their organizational units at an institutional level?

All in all, the purpose of this study is to offer a deeper understanding on what do the classification systems used in bibliometric studies and university rankings represent according universities' organizational units and how can bibliometric indicators ease the interpretation of such fields.

## Material and Methods

In this paper we focus on two Spanish universities as case studies: University of Granada and Pompeu Fabra University. We focus in a single country for two reasons. Firstly, the authors' own expertise and knowledge on the Spanish university system which helps to better interpret the results obtained. Secondly, as Bornmann, Mutz & Daniel (2013) point out, national systems influence the research performance of universities. In this sense, it is of interest to identify differences between universities of a same country. These universities represent two different types of institutions. The former is a historical university with a well-established structure and present in all editions of the Shanghai Ranking and most world-class university rankings. The latter is a small and relatively new university funded in 1990 which has rapidly gained positions in many international rankings converting itself in an interesting success case as pointed out in other studies (Robinson-García et al., 2013b). They are chosen due to their dissimilarity on size, historical background and structure, as one will expect to see some differences regarding the new managerialism of universities and its effect on their institutional organization (Morris, 2002). However, they also have some common grounds as they both belong to the same higher education system. This allows a better interpretation of the results and the effects such structural differences may have on their research performance.

This section is structured as follows. Firstly, we account for the data collection process and the time period used. Secondly, we explain in detail the problems that arise when processing address data and the many limitations it may present. Finally, we describe the methodology employed for analyzing the structure of universities and the indicators proposed for presenting the research profile of each organizational unit and understanding the relation between them and the fields as defined by a given classification system.

*Data collection and processing*

We used the 2006-2010 time period, a fixed four-year citation window and the TS Science Citation Index (SCI), Social Sciences Citation Index (SSCI) and Arts & Humanities Citation Index (A&HCI) as data sources. In order to gather the research output of the universities we used the in-house CWTS version of the TS Web of Science, which identifies universities' output taking into account all possible name variations.

Once the research output of a given university was identified, we delimited our focus only on addresses belonging to the institution under study. Therefore, in a paper published by Pompeu Fabra University in collaboration with London School of Economics we will only consider the affiliation data related to Pompeu Fabra University and omit the one related to the collaborating institution. This way we isolate addresses referring to the institution under study. In Figure 1 we show how this process was followed. Hence, on a first step we identify the address field. As observed in the figure, the record used as an example includes three different addresses delimited by dots, all of them belonging to the University of Granada. This means that none will be discarded. Next, we observe that each address is further divided by organizational units. In step 3, we notice that these are separated by commas and furthermore, that the first unit identifies the major organizational level (in our case, the university) and that the two last ones identify post code and city, and country accordingly. Thus these fields can be automatically removed. Once this has been done, we can identify the rest of the organizational units by dividing the field using commas as separators. In the example used in Figure 1, we identify five different organizational units in step 4. In this case, the address refers to a double affiliation including a school, a faculty, a department and two research groups. This is in fact an interesting example that shows the many inconsistencies one may find when working with address data. On the one hand, it includes two units which belong to the same hierarchical level (school and faculty) which could question if the department belongs to both organizational units or if this address should have been treated as two different addresses. Also, a closer look at the information provided for the two research groups will show that they are in fact the same research group displayed in English and Spanish language.

Although in this example organizational units belonged to four different structural levels, authors may not always indicate all units and may omit the faculty or school for instance, only reflecting departmental information or include only information regarding the highest organizational unit (school and faculty in our example). Indeed, the variability on the information provided by this field varies significantly, ranging from records which do not offer any organizational unit (meaning that they will not be retrieved in this analysis) to records which offer other information which is not always related with their position within the structure of their university. In this sense, authors may indicate as address data the funding agency which supports them, national collaboration networks, research programs or the name of the building in which they work, for instance. Also there are problems when establishing boundaries between units belonging to a university and mixed units with more than one affiliations such as hospitals, research institutes, etc. (for a further discussion the reader is referred to Praal et al. (2013) in which they address the many problems that can arise regarding the assignment of hospitals in the United Kingdom). As in this paper we focus on institutional structure, papers including only the main organizational level (university) were discarded from the analysis. In Figures 2 and 4 we show the distribution of organizational units and the proportion they represent of the total output.

FIGURE 1. Example on the procedure for identifying organizational units within bibliographic records from the Web of Science

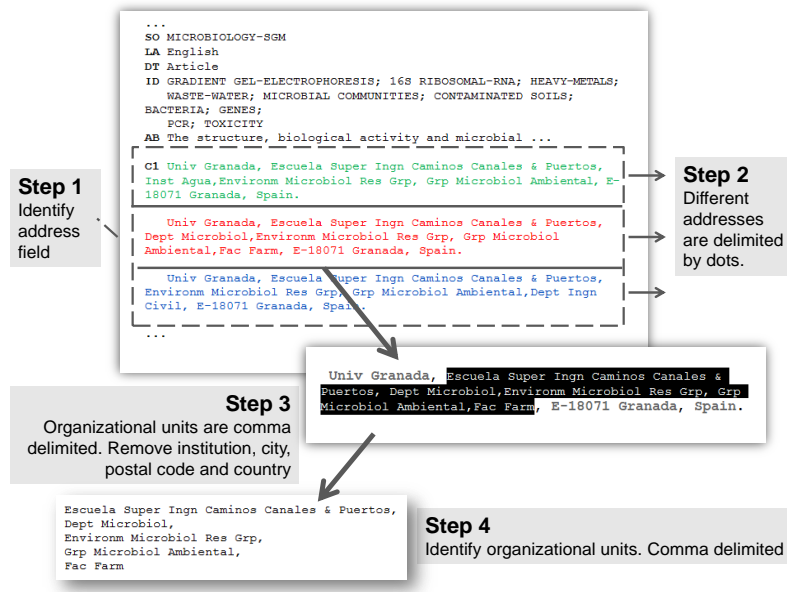

Another relevant issue when dealing with address data has to do with the many normalization problems mainly due to misspellings, use of different languages, name changes, the use of acronyms and errors made by Web of Science; this results in a necessary manual cleaning of data. Regarding the latter problem, we have noted many inconsistencies between the information provided by the authors in publications and the one displayed in the Web of Science. In Table 1 we include the most common denominations out of a total of 41 name variations found for a department belonging to the University of Granada. This data cleaning process was made by checking with the institutional website in order to compare results, although in some cases we found out that certain institutions or units defined by the authors were not included in the institutional layout. Units belonging to parallel structures such as research groups, hospitals, research programs or national collaboration networks were preserved.

TABLE 1. Name variations and number of papers linked to a department from the University of Granada

|  | No PUB* | DEPARTMENT |
|---|---|---|
| **CHOSEN DESIGNATION** | 162 | **DEPT COMP SCI & ARTIFICIAL INTELLIGENCE** |
| **VARIABLES** | 33 | DEPT COMP SCI & AI |
|  | 18 | DEPT CIENCIAS COMPUTAC & IA |
|  | 15 | DEPT CIENCIAS COMPUTAC & INTELIGENCIA ARTIFICIAL |
|  | 9 | DEPT COMP SCI |
|  | 4 | COMP SCI & ARTIFICIAL INTELLIGENCE DEPT |
|  | 4 | DEPT CIENCIAS COMP & INTELIGENCIA ARTIFICIAL |
|  | 4 | DPTO CIENCIAS COMPUTAC & INTELIGENCIA ARTIFICIAL |
|  | 3 | AI |
|  | 3 | DEPT COMP SCI & ARTIFICAL INTELLIGENCE |
|  | 2 | DECSAI |
|  | 2 | DEPT CIENCIAS COMPUTAC |
|  | 2 | DEPT COMPUTAT SCI & AI |
|  | 28 | *OTHER VARIATIONS WITH PUBLICATION FREQUENCE 1* |
| **TOTAL** | **281** |  |

* A department may be included several times in the same paper

*Construction of research profiles and indicators used*

The goal of this study is to understand the relation between fields as constructed in rankings and bibliometric studies, and the structure of universities as defined by their organizational units and offer indicators that can explain such relation. To this aim, we developed an organizational network for each university under study which would allow a general view of its structure according to its research output. Organizational units may co-occur in a document for different reasons. Hence, the following cases may take place: several authors belonging to different departments (in-house collaboration), one author indicating different organizational units all within each hierarchical level (i.e., faculty, department, research group) or one author with double affiliation (i.e., faculty and research center). Therefore, links in our network will define organizational relations between units in its broadest sense.

Such networks are shown in Figures 3 and 5, and they allow us to identify the units which occupy a central or most 'prominent' position in the structure, that is, they have more potential power and influence due to their connections to the rest of the nodes (Borgatti, Mehra, Brass & Labianca, 2009). Here we propose the use of centrality indicators to understand the role of a given organizational unit within the rest of the network. There are different indicators which measure the centrality of nodes; in this paper we use the betweenness indicator. A node will have high betweenness centrality if it appears often in the shortest path that connects any two other nodes. In Figures 3 and 5 the betweenness centrality measure is represented by the size of the nodes. Table 3 and 4 present the research profiles of departments with more than 50 publications in the case of University of Granada and for any organizational unit with more than 50 publications in the case of Pompeu Fabra University. The methodology for the construction of research profiles is based on the work by Calero-Medina & van Leeuwen (2012) and consists on 'breaking down' the output of an organizational unit into subject fields based on a given classification system. This way one can observe the 'interdisciplinarity' of such unit. Finally, we propose to combine the betweenness centrality measure with the Gini Index, as a means to observe how well represented are organizational units by fields of a given classification system. In table 2 we include a list of the indicators employed along with their definition.

TABLE 2. Description of the indicators used.

| Indicator | Acronym | Definition |
|---|---|---|
| Number of publications | P | Publications indexed in the Web of Science citation indexes (SCI, SSCI and A&CI). The considered document types were letters, articles, reviews and proceedings papers. |
| Betweenness Centrality | B | The Betweennes Centrality measure indicates the nodes which appear more often when connecting two other nodes in a network. A node will have high betweenness centrality if it appears often in the shortest path that connects any two other nodes. |
| Gini Coefficient | G | The Gini Coefficient is an inequality indicator which shows the concentration or scattering of distributions. It is commonly used in the field of Economics to analyze the distribution of wealth. In this study we use it to analyze the distribution of an organizational unit's output according to subject fields. Its value ranges from 0 to 1; 0 meaning no concentration and 1 concentration in a single subject field. In this paper we used the formula defined by Deaton (1997). |
| Number of subject categories | No SC | By subject categories we refer to the classification system employed by TS. |
| Number of disciplines | No disc | By disciplines we refer to the classification system employed by the Spanish I-UGR Rankings. Such system is based on TS' subject categories from the SCI and the SSCI, and defines a total of 37 disciplines. The construction of these disciplines is available at http://www.ugr.es/~elrobin/docs/disciplines_I-UGR_Rankings.xlsx. |

Two classification systems were selected according to two possible scenarios in which institutional analyses by fields take place:

1) The TS subject categories. This is the most common classification system used in bibliometric studies. In fact, it is the one employed by Calero-Medina & van Leeuwen (2012), who use it in order to construct inverse research profiles (that is, a breakdown of subject categories into organizational units) as a means to analyze the contribution of different research programs to a given research field.

2) Aggregation of subject categories. A common methodology employed also in bibliometric analyses at a macro-level (i.e., García et al., 2012) and in university rankings by fields (i.e., the 2013 edition of the Leiden Ranking or the Spanish I-UGR Rankings of field and disciplines). In this study we will use the 37 aggregated disciplines defined in the I-UGR Rankings, as this will allow us to discuss the implications of possible discrepancies between fields and organizational units in university rankings.

## Results

*Case 1. University of Granada*

The University of Granada had a total output of 6913 publications for the 2006-2010 time period of which 6337 were finally included in this study. The remaining did not include any information at the organizational unit level. In figure 2 we include a general overview of different types of organizational units used by authors to indicate their affiliation. As observed, the most common information included is the department, which is present in 5514 papers which represent 87.0% of the total output analyzed. Also, this organizational type is the one in which a wider number of units were found with 132 departments. In total nearly half of the total share (48.3%) included information regarding the faculty to which authors belonged, followed by far by papers including information regarding the research center to which authors were affiliated (19.6). The rest of the organizational types account each for less of 10% of the total output. The other organizational type with the largest number of different after departments is others, which is a miscellaneous group in which one may find a wide range of different units such as office, job post or errors in the database such as other universities involved, not following the rationale of the address field as described previously in Figure 1.

FIGURE 2. Percentage of publications and total number of units by type for the University of Granada according to organizational types. Time period 2006-2010

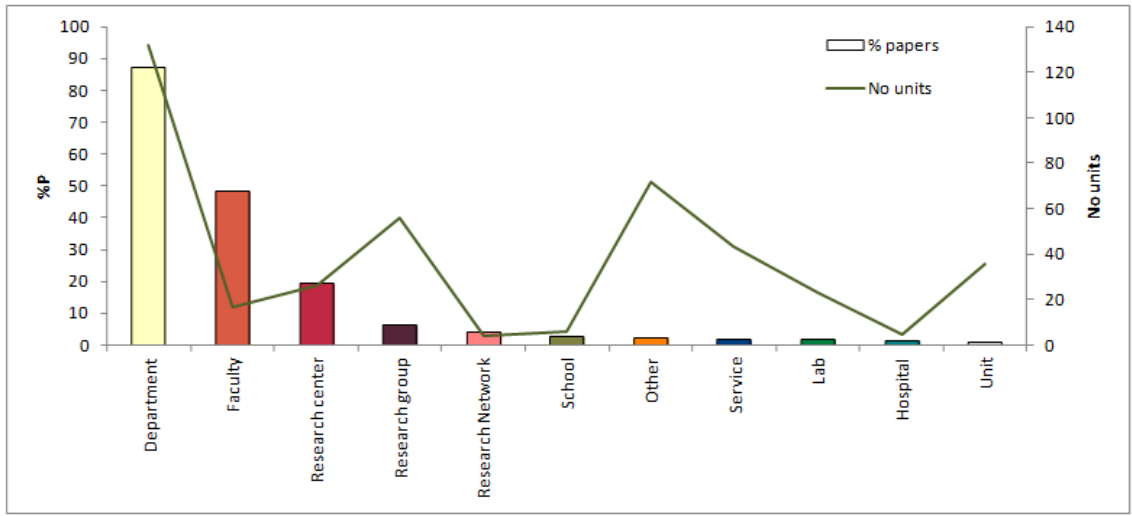

In figure 3 we show the structure of the university according to its organizational units. As it was explained previously the links in the network will define organizational relations between units. As observed, four different components can be found, three small ones and a main component. Units are organized around faculties and departments and occasionally around research centers. These are the main organizational units. This component can be further divided in seven distinct parts. On the upper right we find organizational units related with the fields of Behavioral Sciences and Neuroscience. One of the two main clusters is formed around the Faculty of Sciences FAC SCI) which connects through the Faculty of Pharmacy (FAC PHARM) with the other main cluster representing the Biomedical Sciences and formed by the faculties of Medicine (FAC MED) and Dentistry (FAC DENT). On the upper left we have Engineering and on the lower left, Physics. On the lower right, we find units related with fields from Computer Science which connect with those related with Information Science.

FIGURE 3. Organizational network of the University of Granada according to coupling of organizational units. Map characteristics: Lines: minimum co-occurrence value >5. Isolated nodes have been removed. Colors: Types of organizational units as show in Figure 2. Size: Betweenness values

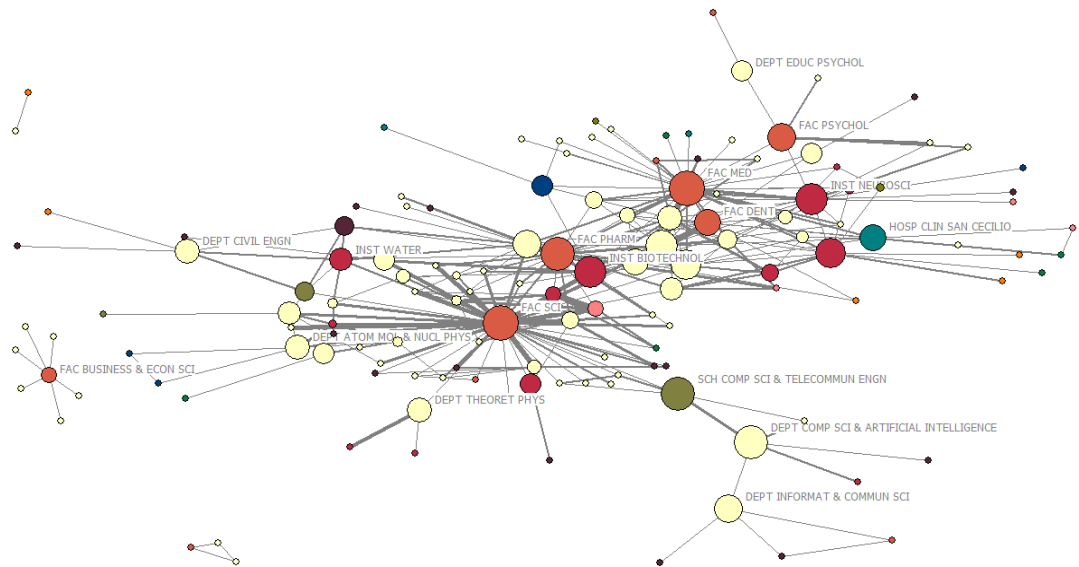

Table 3 includes an overview of the research profiles for all departments of the University of Granada with more than 50 publications in the 2006-2010 time period. For each department we include the total publications, betweenness centrality and their Gini Coefficient and number of subject fields according to each classification system. These three indicators offer valuable information on the research profile of each department and the capability of the classification system to isolate its output. For instance, we observe that the department of Mathematical Analysis (DEPT ANAL MATH) is the one with the highest concentration according to the TS Classification (0.86) with its output distributed among 6 subject categories and all of it included in a single discipline according to the I-UGR classification system. Regarding its importance in the rest of the network, it does not have a central position in terms of being a department that connects units that otherwise will be unconnected. This is the reason why its betweenness centrality is zero. This result goes online with the higher values for the Gini coeficients. The deparment of Mathematical Analysis is very focus on certain fields. Hence, both classification systems can accurately reflect this department's output.

On the other extreme we find the department of Optics (DEPT OPT), which has a 0.62 Gini coefficient distributed among 17 different subject categories according to the TS classification, and a 0.43 Gini coefficient distributed among 10 different disciplines. This means that its output is neglected by subject fields as it is widely distributed. Regarding its role in the institutional structure of the university, it has betweenness value of 126.0. Another different case is the one of departments which, despite concentrating most of their output in certain subject categories, they contribute to many other fields. This occurs with the output of the department of Computer Sciences & Artificial Intelligence (DEPT COMP SCI & ARTIFICIAL INTELLIGENCE) which shows Gini values above 0.7 for both classification systems but performs in 57 different subject categories according to the TS classification and in 20 disciplines according to the I-UGR classifications. Although its output is mainly focused on Computer Sciences, it also performs in many other areas, this is also observed by its relation within the network where its betweennes centrality has a value of 735.0. Other examples of this case can be observed on the department of Statistics & Operational Research (DEPT STAT & OPERAT RES) and the department of Applied Physics (DEPT APPL PHYS). In fact, these three departments are the most productive ones leading to suggest that the more output produced by a department, the larger the contribution to other fields may be.

TABLE 3. Bibliometric indicators and research profiles of departments from University of Granada with >50 publications for the 2006-2010 time period and their output distribution according to two classification systems: TS subject categories and I-UGR disciplines

| Department | P | B | TS CLASSIFICATION | | I-UGR CLASSIFICATION | |
|---|---|---|---|---|---|---|
| | | | G | No SC | G | No disc |
| DEPT COMP SCI & ARTIFICIAL INTELLIGENCE | 281 | 735.00 | 0.76 | 57 | 0.84* | 20 |
| DEPT APPL PHYS | 266 | 34.02 | 0.63 | 62 | 0.75 | 21 |
| DEPT STAT & OPERAT RES | 239 | 682.72 | 0.62 | 75 | 0.62 | 23 |
| DEPT ZOOL | 231 | 0.00 | 0.67 | 36 | 0.74 | 14 |
| DEPT INORGAN CHEM | 225 | 0.00 | 0.69 | 34 | 0.82* | 14 |
| DEPT ATOM MOL & NUCL PHYS | 223 | 250.67 | 0.64 | 26 | 0.67 | 6 |
| DEPT THEORET PHYS | 221 | 251.00 | 0.77 | 12 | 0.77 | 3 |
| DEPT ANALYT CHEM | 218 | 5.00 | 0.77 | 41 | 0.79 | 13 |
| DEPT PHYSIOL | 203 | 235.60 | 0.64 | 44 | 0.71 | 15 |
| DEPT STOMATOL | 180 | 339.98 | 0.80* | 31 | 0.79 | 9 |
| DEPT APPL MATH | 176 | 136.50 | 0.76 | 37 | 0.67 | 12 |
| DEPT MINERAL & PETROL | 175 | 4.01 | 0.66 | 46 | 0.77 | 13 |
| DEPT MICROBIOL | 165 | 405.64 | 0.69 | 45 | 0.62 | 15 |
| DEPT STRATIG & PALEONTOL | 164 | 0.50 | 0.73 | 31 | 0.84* | 9 |
| DEPT GEOMETR & TOPOL | 140 | 0.00 | 0.77 | 9 | 1.00* | 1 |
| DEPT MATH ANAL | 136 | 0.00 | 0.86* | 6 | 1.00* | 1 |
| DEPT EXPT PSYCHOL | 124 | 0.00 | 0.68 | 36 | 0.81* | 12 |
| DEPT ORGAN CHEM | 118 | 126.00 | 0.70 | 26 | 0.87* | 12 |
| DEPT CHEM ENGN | 109 | 0.00 | 0.61 | 27 | 0.61 | 14 |
| DEPT GEODYNAM | 109 | 0.00 | 0.68 | 25 | 0.87* | 8 |
| DEPT COMP ARCHITECTURE & TECHNOL | 97 | 0.00 | 0.62 | 35 | 0.78 | 14 |
| DEPT PHARMACOL | 97 | 60.63 | 0.58 | 38 | 0.70 | 14 |
| DEPT ECOL | 96 | 3.83 | 0.61 | 23 | 0.63 | 9 |
| DEPT CIVIL ENGN | 91 | 251.00 | 0.61 | 42 | 0.67 | 18 |
| DEPT ELECTROMAGNET & PHYS MATTER | 88 | 4.50 | 0.57 | 29 | 0.56 | 9 |
| DEPT PHARM & PHARMACEUT TECHNOL | 82 | 0.00 | 0.60 | 24 | 0.71 | 9 |
| DEPT PLANT PHYSIOL | 81 | 0.00 | 0.62 | 28 | 0.65 | 12 |
| DEPT OPT | 77 | 126.00 | 0.62 | 17 | **0.43** | 10 |
| DEPT PERSONAL ASSESSMENT & PSYCHOL TREATMENT | 75 | 0.00 | 0.59 | 35 | 0.67 | 9 |
| DEPT PHYS CHEM | 75 | 0.00 | 0.64 | 20 | 0.70 | 8 |
| DEPT GENET | 74 | 7.43 | 0.56 | 25 | 0.65 | 9 |
| DEPT NUTR & BROMATOL | 74 | 0.00 | 0.67 | 16 | 0.59 | 10 |
| DEPT ELECT & COMP TECHNOL | 72 | 0.50 | 0.70 | 26 | 0.72 | 8 |
| DEPT BIOCHEM & MOL BIOL | 69 | 536.69 | 0.52 | 37 | 0.70 | 11 |
| DEPT ALGEBRA | 67 | 0.00 | 0.74 | 3 | 1.00* | 1 |
| DEPT BOT | 66 | 2.45 | 0.54 | 32 | 0.65 | 14 |
| DEPT INFORMAT & COMMUN SCI | 66 | 374.00 | 0.79 | 13 | 0.80 | 6 |
| DEPT ANAT PATHOL & HIST CIENCIA | 58 | 19.22 | **0.43** | 33 | 0.64 | 11 |
| DEPT LANGUAGES & COMP SYST | 58 | 0.00 | 0.57 | 29 | 0.72 | 13 |
| DEPT SIGNAL THEORY TELEMAT & COMMUN | 56 | 0.00 | 0.53 | 29 | 0.61 | 13 |
| DEPT BIOCHEM & MOL BIOL 2 | 54 | 194.04 | 0.51 | 25 | 0.62 | 9 |

In bold when G <0.5; with an * when G>0.8

## Case 2. Pompeu Fabra University

Pompeu Fabra University had a total output of 2480 publications for the 2006-2010 time period of which 1760 were finally included in this study. The remaining publications did not include any information at the organizational unit level. As it occurred with University of Granada, department is the most frequent organizational type included in its publications, although in this case its share drops to 63.5% (1117 publications). In fact, more variability is found in the organizational types adopted by authors at this university. Also the order of the most present types varies and research centers are the second most frequent choice accounting for 34.3% of the total share followed by units which represent 21.2% of the output. The faculty name is rarely used and it represents 3.5% of the total share, that is, 61 publications. Regarding the number of units, the structure of the university also differs from the first case study. There are less departments (18), and faculties (7) and the greatest number of units can be found in the miscellaneous group 'Other' (47), followed by research centers (34) and then units (27) and research groups (21).

FIGURE 4. Percentage of publications and total number of units by type for the Pompeu Fabra University according to organizational types. Time period 2006-2010

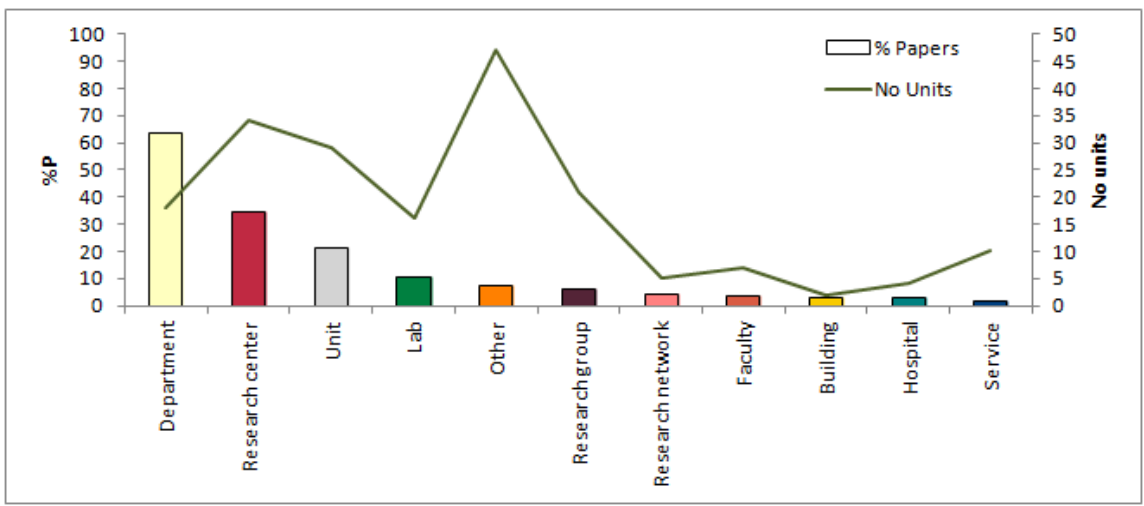

In figure 5 we show the structure of the university according to its organizational units. In this case, units are organized around departments, research centers and occasionally around units. Contrarily to Granada, and despite having seven faculties and a school, these organizational units are almost absent in the address data offered by researchers from Pompeu Fabra. The network is formed by a single component. Also, we observe that, except the lower left of the figure, most of the organizational units are related with fields from the Biomedical Sciences, displaying a highly specialized university. The organizational units displayed on the lower left are related with fields from the Social Sciences and Computer Science.

FIGURE 5. Organizational network of Pompeu Fabra University according to coupling of organizational units. Map characteristics: Lines: minimum co-occurrence value >3. Isolated nodes have been removed. Colors: Types of organizational units as show in Figure 2. Size: Betweenness values

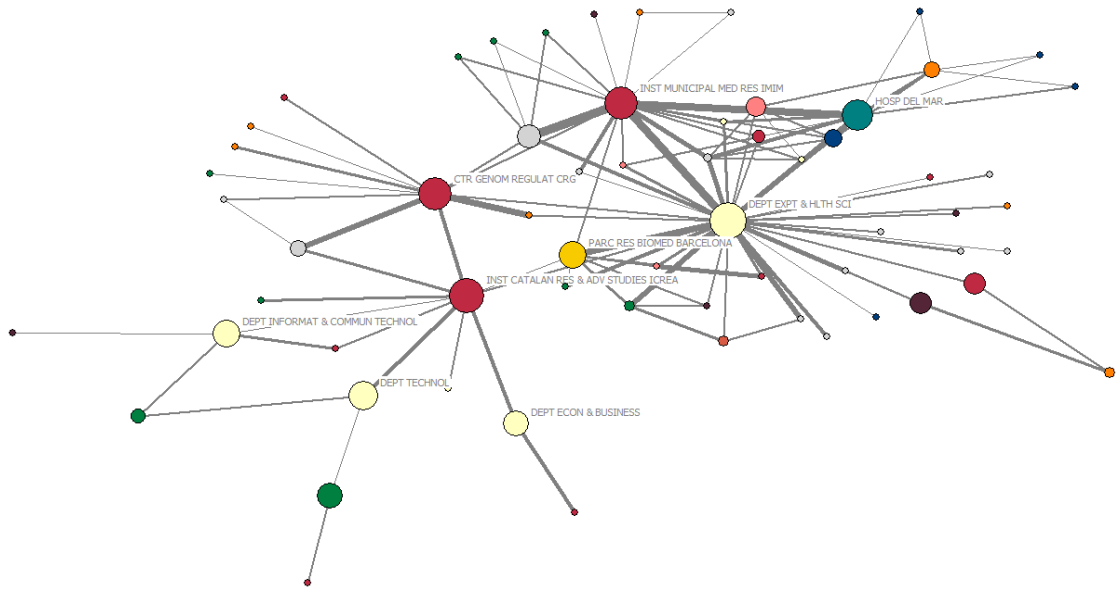

Due to the larger distribution of organizational types along with lower output figures, in this case study we have considered all organizational units with more than 50 publications and not only departments. The research profiles of each of them along with some bibliometric indicators are shown in table 4. In fact, only four departments are above such threshold, the most productive of them, the department of Experimental & Health Sciences (DEPT EXPT & HLTH SCI) accounting for 34.0% of the total share of Pompeu Fabra University. Also we find other institutions included which do not actually belong to this university. It is the case of the Institut Municipal D'Investigacions Mèdiques (INST MUNICIPAL MED RES IMIM) which is a mixed institution belonging Hospital del Mar but whose staff is affiliated to various institutions such as Pompeu Fabra University, Autonomous University of Barcelona or the Centre of Genomic Regulation. This institution along with the latter (CTR GENOM REGULAT CRG), are both mixed research centers with staff from different Catalan universities. They all belong to the Barcelona Biomedical Research Park (PARC RES BIOMED BARCELONA), where along with others, these research centers are located.

TABLE 4. Bibliometric indicators and research profiles of organizational from Pompeu Fabra University with >50 publications for the 2006-2010 time period and their output distribution according to two classification systems: TS subject categories and I-UGR disciplines

| Department | P | B | TS CLASSIFICATION | | I-UGR CLASSIFICATION | |
| --- | --- | --- | --- | --- | --- | --- |
| | | | G | No SC | G | No disc |
| DEPT EXPT & HLTH SCI | 599 | 1257.28 | 0.67 | 88 | 0.70 | 21 |
| DEPT ECON & BUSINESS | 206 | 61.00 | 0.67 | 53 | 0.71 | 17 |
| INST MUNICIPAL MED RES IMIM | 166 | 362.31 | 0.59 | 55 | 0.64 | 14 |
| CTR GENOM REGULAT CRG | 155 | 361.48 | 0.67 | 33 | 0.62 | 10 |
| INST CATALAN RES & ADV STUDIES ICREA | 122 | 618.83 | **0.45** | 51 | **0.34** | 14 |
| DEPT TECHNOL | 119 | 148.00 | 0.59 | 44 | 0.70 | 15 |
| DEPT INFORMAT & COMMUN TECHNOL | 96 | 90.00 | 0.54 | 43 | 0.66 | 13 |
| UNIT BIOMED INFORMAT GRIB | 88 | 56.90 | 0.60 | 35 | 0.55 | 10 |
| UNIT EVOLUT BIOL | 64 | 0.00 | 0.64 | 17 | 0.65 | 8 |
| LAB NEUROPHARM | 54 | 1.00 | 0.66 | 12 | 0.56 | 6 |
| PARC RES BIOMED BARCELONA | 52 | 73.87 | 0.50 | 35 | 0.57 | 13 |
| HOSP DEL MAR | 51 | 195.96 | 0.50 | 27 | 0.73 | 8 |

In bold when G <0.5

According to their research profile, none of these organizational units show values above 0.8 on their Gini Coefficient for any of the classifications used, distributing their research output in a wide range of subject fields. In fact, we observe that two units show Gini values under 0.5 according to the TS classification: INST CATALAN RES & ADV STUDIES ICREA (0.45) and HOSP DEL MAR (0.50). On the first case, this is quite normal as this institution is a multidisciplinary agency from the regional government focused on recruiting international researchers and integrating them in research centers and universities located within Catalonia. The second case is a hospital and it responds reasonably well. As observed, although it shows a low Gini coefficient when using the TS classification, it value raises up to 0.730 when using the I-UGR classification. Finally, we find that the betweenness centrality of the organizational present in table 4 are much higher than in the previous case, with some exceptions, showcasing a much more integrating and multidisciplinary structure of university.

## Discussion

In this paper we highlight the problems that may arise when interpreting university rankings by fields as these are commonly mistaken with organizational units within the structure of universities. Also, we propose the use of the Gini Index and the betweenness centrality measure as a means to understand how well are different organizational units represented by the field classification systems employed in bibliometric studies and rankings by fields. For this purpose we focus on two Spanish universities as case studies, Granada and Pompeu Fabra, which reflect two different types of institutions. Granada represents a historical university with a well-established structure while Pompeu Fabra represents a young and dynamic institution with an outstanding research performance. Then we develop a research profile for each department/organizational unit according to two different classification systems (TS subject categories and I-UGR Rankings disciplines) in order to showcase the discrepancies between the organizational units and the fields of each classification system.

Before discussing our results, we must emphasize on the implicit problems that working with addresses brings to any bibliometric analysis when adopting a *top down* approach. As it has been acknowledged elsewhere (Waltman et al., 2012), identifying institutions based on the address field of TS Web of Science means to inevitably assume some errors on the data retrieval of academic institutions. In this study we have shown that the problem may be even worse when deepening on organizational units within universities, leading to the need of manual data cleaning. Although many efforts have been done on standardizing and automatically retrieving address data (a good overview is included in Cuxac, Lamirel & Bonvallot, 2013), still the problem remains unsolved, especially at a large scale where first-hand institutional information is needed in order to verify the data provided from the database (van Leeuwen, 2007).

If solved, bibliometric analyses and rankings could greatly benefit from this kind of approach. As we see in Figures 3 and 5, no only it is possible to understand and analyze the structure of universities according to address data, but also organizational units group themselves according to fields. The structure and size of universities varies significantly due to the managerial changes that have taken place at the end of the 20$^{th}$ century, influenced by different socio-economic factors such as the expansion of higher education and the demand for return on investment, largely exemplified by the organizational forms defined by Gibbons et al. (1994). These changes prevent from the use of address data to construct fields as proposed by De Bruin & Moed (1993). In this sense, we find notable differences between the structure of each university, especially on the loss of importance faculties as an organizational form play in Pompeu Fabra and an increasing importance of departments along with research centers as the main joints in the university. In fact, departments in this university do not seem to be any longer the basic administrative unit and are replaced in such function by research groups, labs and units. Also, as observed in table 4, we find that many organizational units behave as expected according to mode 2 and are 'based outside the university and its traditional disciplinary structure' (Morris, 2002). On the other side we find that Granada still obeys to such a disciplinary structure and in fact, higher values of concentration can be observed when developing research profiles for each department (table 3).

Finally, we must emphasize the clarity with which the Gini coefficient along with the betweenness centrality value and the number of subject fields in which each organizational unit performs, reflect the levels of discrepancy between organizational units and the classification system used. Hence, we highlight the importance of university rankings by fields to provide clear instructions on the classification system used along with the necessary tools so that such profiles can be easily developed by third parties.

## Concluding remarks and further research

The need for accurate and reliable data is a key issue when developing bibliometric tools and studies, and is in fact, one of the main weaknesses of university rankings (van Raan, 2005). The main problem is located in the use of bibliographic data which was not originally conceived to be used for bibliometric purposes and hence lacks of the standardization needed for this type of analyses. The rise of rankings has also raised other more fundamental questions which are still unsolved and which should be addressed before attempting at any institutional comparisons, such as what is a university? What does it mean to belong to a university? How should mixed institutions with more than on affiliation be treated? It seems that the only reasonable way to certify the accuracy and reliability of such studies is having some kind of output verification from the institutions involved. In this sense, further research is needed on analyzing the congruence between the affiliation of authors and the one they indicate in the address field, however, for this type of study to take place we would need internal information from the institution which is normally unavailable. But there is much at stake and the increasing need to offer global products which show where institutions stand at an international level leaves little choice but to assume these problems.

Rankings by fields intend to take into account the disciplinary focus of universities. But because they lack the proper information regarding the structure of universities, they are obliged to use other classification systems based on universities' research output. However, this misleads the user of rankings who expects to see on those fields a reflection of the structure of their universities. We have noted that for rankings offering a wide range of league tables by field, not only the media but also research managers tend to confound these with organizational units. Believing that a ranking on a specific field to be a ranking of faculties for instance, or not understanding why a given university can gain certain positions in fields which are not reflected on their institutional structure. In this paper we provide three measures for representing the research profiles of organizational units as a possible solution to show the levels of discrepancy of the fields offered by rankings and the structure of universities.

## Acknowledgments

Thanks are due to the two anonymous referees for their constructive suggestions. The authors would also like to thank Thed N. van Leeuwen and Daniel Torres-Salinas for their helpful comments on previous versions of this paper. Nicolás Robinson-García is currently supported by a FPU grant from the Spanish Ministerio de Economía y Competitividad.